\documentstyle[12pt]{article}
\newlength{\extraspace}
\newlength{\extraspaces}
\newcommand{\be}{\begin{equation}
\addtolength{\abovedisplayskip}{\extraspaces}
\addtolength{\belowdisplayskip}{\extraspaces}
\addtolength{\abovedisplayshortskip}{\extraspace}
\addtolength{\belowdisplayshortskip}{\extraspace}}
\newcommand{\ee}{\end{equation}}
\newcommand{\ba}{\begin{eqnarray}
\addtolength{\abovedisplayskip}{\extraspaces}
\addtolength{\belowdisplayskip}{\extraspaces}
\addtolength{\abovedisplayshortskip}{\extraspace}
\addtolength{\belowdisplayshortskip}{\extraspace}}
\newcommand{\ea}{\end{eqnarray}}
\newcommand{\nonu}{\nonumber \\[.5mm]}
\newcommand{\A}{&\!\!\!}

\begin{document}
\addtolength{\baselineskip}{.7mm}
\begin{flushright}
STUPP-96-145 \\ March, 1996
\end{flushright}
\vspace{.6cm}
\begin{center}
{\large{\bf{Supersymmetry algebra 
            in $N = 1$ chiral supergravity}}} \\[20mm]
{\sc Motomu Tsuda and Takeshi Shirafuji} \\[12mm]
{\it Physics Department, Saitama University \\[2mm]
Urawa, Saitama 338, Japan} \\[20mm]
{\bf Abstract}\\[10mm]
{\parbox{13cm}{\hspace{5mm}
We consider the supersymmetry (SUSY) transformations 
in the chiral Lagrangian 
for $N = 1$ supergravity (SUGRA) with the complex tetrad 
following the method used in the usual $N = 1$ SUGRA, 
and present the explicit form of the SUSY 
trasformations in the first-order form. 
The SUSY transformations are generated 
by two independent Majorana spinor parameters, 
which are apparently different from the constrained 
parameters employed in the method of the 2-form gravity. 
We also calculate the commutator algebra of 
the SUSY transformations on-shell.}} 
\end{center}
\vfill

\newpage
%

Ashtekar's canonical formulation of general relativity 
was extended to $N = 1$ supergravity (SUGRA) 
introducing the right- and left-handed supersymmetry (SUSY) 
transforamtions \cite{AAN,JJ}. 
The first-order formulation and its extension 
to $N = 2$ SUGRA were made using the 2-form gravity 
\cite{CDJ,KS}. 
In this formulation, however, the SUSY transformation 
parameters are constrained. 
The purpose of this brief report is to reconsider 
the SUSY transformations in the $N = 1$ chiral SUGRA, 
following as closely as possible 
the method originally employed 
in the usual $N = 1$ SUGRA \cite{FN,DZ}. 

The $N = 1$ chiral SUGRA has characteristic features 
to be contrasted with the usual $N = 1$ SUGRA: 
Firstly a complex tetrad field is introduced, 
and secondly spin-3/2 fields $\psi_{\mu}$ 
and $\overline \psi_{\mu}$ are assumed to be independent 
of each other. 
The motivations of taking such chiral Lagrangian 
as analytic in complex field variables 
are (a) to evade a consistency problem for matter field equations 
when more than two spin-3/2 fields are coupled, 
\footnote{\ If the tetrad is real and the self-dual 
connection satisfies its equation of motion, 
the chiral Lagrangian including more than two spin-3/2 
fields becomes complex, 
and its imaginary part gives an additional equation 
for spin-3/2 fields which gives rise to inconsistency 
\cite{TSX}.} 
and (b) to construct the SUSY transformations 
compatible with the complex tetrad. 
We present the explicit form of the SUSY 
transformations in the first-order form. 
The present formulation has the merit that 
the SUSY transformation parameters are not constrained at all 
in contrast with the method of the 2-form gravity. 
We also calculate the commutator algebra 
of the SUSY transformations on-shell. 

We start with the chiral Lagrangian density 
for $N = 1$ SUGRA, 
\be
{\cal L}^{(+)} = {\cal L}^{(+)}_G + {\cal L}^{(+)}_{RS}. 
\label{L+}
\ee
The independent variables in ${\cal L}^{(+)}$ are a {\it complex} 
tetrad $e_{\mu}^i$, a self-dual connection $A^{(+)}_{ij \mu} 
= A^{(+)}_{[ij] \mu}$ which satisfies 
$(1/2){\epsilon_{ij}} \! ^{kl} A^{(+)}_{kl \mu} 
= i A^{(+)}_{ij \mu}$, 
and two {\it independent} (Majorana) Rarita-Schwinger 
fields $\psi_{R \mu}(= (1/2)(1 + \gamma_5) \psi_{\mu})$ 
and $\overline{\tilde \psi}_{R \mu}$. 
\footnote{\ Greek letters $\mu, \nu, \cdots$ are 
space-time indices, and Latin letters {\it i, j,} $\cdots$ 
are local Lorentz indices. 
We denote the Minkowski metric 
by $\eta_{ij} =$ diag$(-1, +1, +1, +1)$. 
The totally antisymmetric tensor $\epsilon_{ijkl}$ 
is normalized as $\epsilon_{0123} = +1$. 
The antisymmetrization of a tensor with respect to $i$ 
and $j$ is denoted 
by $A_{[i \mid \cdots \mid j]} 
:= (1/2)(A_{i \cdots j} - A_{j \cdots i})$.} 
The chiral gravitational Lagrangian density, 
${\cal L}^{(+)}_G$, constructed from the complex tetrad 
and the self-dual connection is 
\be
{\cal L}^{(+)}_G = -{i \over 2} e \ \epsilon^{\mu \nu \rho \sigma} 
   e_{\mu}^i e_{\nu}^j R^{(+)}_{ij \rho \sigma}, 
\label{LG+}
\ee
where the unit with $8 \pi G = c = 1$ is used, 
$e$ denotes ${\rm det}(e^i_{\mu})$ 
and the curvature of self-dual connection 
${R^{(+)ij}}_{\mu \nu}$ is 
\be
{R^{(+)ij}}_{\mu \nu} := 2(\partial_{[\mu} {A^{(+)ij}}_{\nu]} 
             + {A^{(+)i}}_{k [\mu} {A^{(+)kj}}_{\nu]}). 
\label{curv+}
\ee
The chiral Lagrangian density of (Majorana) Rarita-Schwinger fields, 
${\cal L}^{(+)}_{RS}$, is 
\be
{\cal L}^{(+)}_{RS} = - e \ \epsilon^{\mu \nu \rho \sigma} 
                     \overline{\tilde \psi}_{R \mu} \gamma_\rho 
                     D^{(+)}_\sigma \psi_{R \nu}, 
\label{LRSE+}
\ee
where $D^{(+)}_\mu$ denotes 
the covariant derivative with respect to $A^{(+)}_{ij \mu}$: 
\be
D^{(+)}_\mu := \partial_\mu + {i \over 2} A^{(+)}_{ij \mu} S^{ij} 
\ee
with $S_{ij}$ being the Lorentz generator. 
\footnote{\ In our convention $S_{ij} = {i \over 4}[\gamma_i, \gamma_j]$ 
and $\{ \gamma_i, \gamma_j \} = - 2 \eta_{ij}$.} 

The field equations derived from the ${\cal L}^{(+)}$ of (\ref{L+}) 
are slightly different from the usual $N = 1$ SUGRA. 
Varying ${\cal L}^{(+)}$ with respect to $A^{(+)}_{ij \mu}$ 
and solving the equation for $A^{(+)}_{ij \mu}$ yield 
\be
A^{(+)}_{ij \mu} = A^{(+)}_{ij \mu}(e) + K^{(+)}_{ij \mu}, 
\label{solA+}
\ee
where $A^{(+)}_{ij \mu}(e)$ is the self-dual part 
of the Ricci rotation coefficients $A_{ij \mu}(e)$, 
while $K^{(+)}_{ij \mu}$ is that of $K_{ij \mu}$ given by 
\be
K_{ij \mu} := {i \over 2} (e^{\rho}_i e^{\sigma}_j e_{\mu}^k 
             \ \overline{\tilde \psi}_{R [\rho} 
                     \gamma_{\mid k \mid} \psi_{R \sigma]} 
    + e^{\rho}_i 
             \ \overline{\tilde \psi}_{R [\rho} 
                     \gamma_{\mid j \mid} \psi_{R \mu]} 
    - e^{\rho}_j 
             \ \overline{\tilde \psi}_{R [\rho} 
                     \gamma_{\mid i \mid} \psi_{R \mu]}). 
\label{K-right}
\ee
>From (\ref{K-right}) we obtain 
\be
{T^i}_{\mu \nu} := 2 {K^i}_{[\mu \nu]} 
                 = -i \overline{\tilde \psi}_{R [\mu} 
                      \gamma^i \psi_{R \nu]}. 
\label{torsion}
\ee
Varying ${\cal L}^{(+)}$ with respect to $e_{\mu}^i$, 
$\overline{\tilde \psi}_{R \mu}$ and $\psi_{R \mu}$ yields 
\ba
\A \A \epsilon^{\mu \nu \rho \sigma}(e_{\nu}^j R^{(+)}_{ij \rho \sigma} 
      + i \overline{\tilde \psi}_{R \rho} \gamma_i 
      D^{(+)}_{\sigma} \psi_{R \nu}) = 0, 
\label{eq-e} \\
\A \A \epsilon^{\mu \nu \rho \sigma} \gamma_{\rho} 
      D^{(+)}_{\sigma} \psi_{R \nu} = 0, 
\label{eq-RS} \\
\A \A \epsilon^{\mu \nu \rho \sigma} D^{(+)}_{\sigma} 
      (\gamma_{\rho} \tilde \psi_{L \nu}) = 0, 
\label{eq-tRS}
\ea
respectively. 
If the tetrad is real together with $\overline{\tilde \psi}_{R \mu} 
= \overline \psi_{R \mu}$ and if the self-dual connection satisfies 
its equation of motion, 
the field equations of (\ref{eq-e}) to (\ref{eq-tRS}) 
are equivalent to those of the usual $N = 1$ SUGRA. 

It is possible to establish the right- and left-handed 
SUSY transformations in the ${\cal L}^{(+)}$ of (\ref{L+}) 
as in the case of the real tetrad \cite{AAN,JJ}. 
Since $\psi_{R \mu}$ and $\overline{\tilde \psi}_{R \mu}$ 
in (\ref{LRSE+}) are independent of each other, 
we need two anticommuting Majorana spinor parameters 
$\alpha$ and $\tilde \alpha$ which generate 
the SUSY transformations. 
The chiral Lagrangian density ${\cal L}^{(+)}$ 
is invariant under the right-handed SUSY transformations 
generated by $\alpha$, 
\ba
\A \A \delta_R \psi_{R \mu} = 2 D^{(+)}_{\mu} \alpha_R 
      \ \ \ \ \ \delta_R \tilde \psi_{L \mu} = 0, 
\label{psi-psi} \\
\A \A \delta_R e_{\mu}^i 
      = -i \overline{\tilde \psi}_{R \mu} \gamma^i \alpha_R, 
\label{psi-e} \\
\A \A \delta_R A^{(+)}_{ij \mu} = 0, 
\label{psi-A+}
\ea
and also under the left-handed SUSY transformations 
generated by $\tilde \alpha$, 
\ba
\A \A \delta_L \psi_{R \mu} = 0 
      \ \ \ \ \ \delta_L \tilde \psi_{L \mu} 
      = 2 D^{(-)}_{\mu} \tilde \alpha_L, 
\label{tpsi-psi} \\
\A \A \delta_L e_{\mu}^i 
      = -i \overline \psi_{L \mu} \gamma^i \tilde \alpha_L, 
\label{tpsi-e} \\
\A \A \delta_L A^{(+)}_{ij \mu} 
      = -{1 \over 2} \left\{ 
      (\tilde B_{(R)\mu ij} 
             - e_{\mu [i} \tilde B_{(R)}^{\ \ m}{}_{\mid m \mid j]}) 
      -{i \over 2} {\epsilon_{ij}}^{kl} 
      (\tilde B_{(R)\mu kl} 
             - e_{\mu k} \tilde B_{(R)}^{\ \ m}{}_{ml}) \right\} 
\label{tpsi-A+}
\ea
with 
\be
\tilde B_{(R)}^{\lambda \mu \nu} 
:= \epsilon^{\mu \nu \rho \sigma} 
   \overline{\tilde \alpha}_R \gamma^{\lambda} 
   D^{(+)}_{\rho} \psi_{R \sigma}. 
\ee
Here we put a tilde on $B_{(R)}^{\lambda \mu \nu}$ 
because the parameter $\tilde \alpha$ is used. 
In (\ref{tpsi-psi}) of the left-handed SUSY transformations, 
$D^{(-)}_{\mu}$ denotes the covariant derivative 
with respect to antiself-dual connection, $A^{(-)}_{ij \mu}$, 
and we assume that $A^{(-)}_{ij \mu}$ is the solution 
derived from the ``unphysical'' Lagrangian density, 
${\cal L}^{(-)}$: 
\be
{\cal L}^{(-)} = {i \over 2} e \ \epsilon^{\mu \nu \rho \sigma} 
                 e_{\mu}^i e_{\nu}^j R^{(-)}_{ij \rho \sigma} 
                 + e \ \epsilon^{\mu \nu \rho \sigma} 
                 \overline \psi_{L \mu} \gamma_\rho 
                 D^{(-)}_\sigma \tilde \psi_{L \nu}, 
\label{L-}
\ee
where ${R^{(-)ij}}_{\mu \nu}$ is the curvature 
of antiself-dual connection. 
If the tetrad is real and $\overline{\tilde \psi}_{R \mu} 
= \overline \psi_{R \mu}$, 
the ${\cal L}^{(-)}$ of (\ref{L-}) 
becomes just the complex conjugate of the ${\cal L}^{(+)}$ 
of (\ref{L+}). 

In the SUSY transformations in ${\cal L}^{(+)}$, 
the self-dual connection $A^{(+)}_{ij \mu}$ 
is one of the independent variables, 
while the antiself-dual connection $A^{(-)}_{ij \mu}$ 
is the solution derived from ${\cal L}^{(-)}$. 
If we consider the SUSY transformations in ${\cal L}^{(-)}$, 
however, the role of $A^{(+)}_{ij \mu}$ and $A^{(-)}_{ij \mu}$ 
is exchanged each other. 
Indeed, the ${\cal L}^{(-)}$ of (\ref{L-}) is invariant 
under the SUSY transformations 
which has the same form as in ${\cal L}^{(+)}$, if we take 
\be
\delta_R A^{(-)}_{ij \mu} 
= {1 \over 2} \left\{ (B_{(L)\mu ij} 
            - e_{\mu [i} B_{(L)}^{\ \ m}{}_{\mid m \mid j]}) 
  +{i \over 2} {\epsilon_{ij}}^{kl} 
  (B_{(L)\mu kl} - e_{\mu k} B_{(L)}^{\ \ m}{}_{ml}) \right\} 
\label{deltA-}
\ee
with 
\be
B_{(L)}^{\lambda \mu \nu} 
:= \epsilon^{\mu \nu \rho \sigma} 
   \overline \alpha_L \gamma^{\lambda} 
   D^{(-)}_{\rho} \tilde \psi_{L \sigma}, 
\ee
for the right-handed SUSY transformations, and 
\be
\delta_L A^{(-)}_{ij \mu} = 0 
\label{tdeltA-}
\ee
for the left-handed SUSY transformations. 
But in this case, we assume that 
the self-dual connection $A^{(+)}_{ij \mu}$ 
is the solution derived 
from the chiral Lagrangian ${\cal L}^{(+)}$. 

If the tetrad is real and $\overline{\tilde \psi}_{R \mu} 
= \overline \psi_{R \mu}$, 
the left-handed SUSY transformation of the self-dual connection, 
(\ref{tpsi-A+}), can be written as 
\ba
\delta_L A^{(+)}_{ij \mu} 
({\rm 1st \! \! - \! \! order}) 
\A = \A {\rm self \! \! - \! \! dual \ part \ of} \nonu
\A \A {1 \over 2} \{ \delta A_{ij \mu} 
         ({\rm 1st \! \! - \! \! order}) 
         \mid_{N = 1 \ {\rm SUGRA}} \nonu
\A \A - ({\epsilon_{ij}}^{\rho \sigma} \overline \alpha 
   \gamma_{\mu} D_{\rho} \psi_{\sigma} 
   + e_{\mu [i} {\epsilon_{j]}}^{\lambda \rho \sigma} 
   \overline \alpha \gamma_{\lambda} D_{\rho} 
   \psi_{\sigma}) \}. 
\label{real-delA}
\ea
This form of (\ref{real-delA}) 
does not agree with the self-dual part 
of the first-order transformation 
of the connection in the usual $N = 1$ SUGRA. 
In the second-order formulation, however, 
the situation is changed. 
Indeed, if we use the equation for the self-dual connection, 
the spin-3/2 field equation 
$\epsilon^{\mu \nu \rho \sigma} \gamma_{\rho} 
D_{\sigma} \psi_{\nu} = 0$ 
and its variant forms \cite{Nieu} 
\ba
\A \A D_{[\mu} \psi_{\nu]} 
      +{i \over 2} {\epsilon_{\mu \nu}}^{\rho \sigma} \gamma_5 
      D_{\rho} \psi_{\sigma} = 0, \\
\A \A \gamma_{\mu} D_{[\nu} \psi_{\lambda]} 
      + \gamma_{\nu} D_{[\lambda} \psi_{\mu]} 
      + \gamma_{\lambda} D_{[\mu} \psi_{\nu]} = 0, 
\ea
then we can show that (\ref{real-delA}) becomes 
\ba
\delta_L A^{(+)}_{ij \mu} ({\rm 2nd \! \! - \! \! order}) 
\A = \A {\rm self \! \! - \! \! dual \ part \ of} \nonu
\A \A \delta A_{ij \mu} ({\rm 2nd \! \! - \! \! order}) 
      \mid_{N = 1 \ {\rm SUGRA}}. 
\ea

On the other hand, 
the transformations of (\ref{psi-A+}) and (\ref{tpsi-A+}) 
seem to be quite different from the SUSY transformations 
based on the 2-form gravity \cite{CDJ,KS}: 
In fact, in the 2-form gravity, 
the SUSY transformation parameters with constraint are used, 
and the transformations of the connection 
do not include the covariant derivative. 

For the purpose of calculating the SUSY algebra, 
let us write the other equivalent forms 
of the spin-3/2 field equations 
(\ref{eq-RS}) and (\ref{eq-tRS}) explicitly. 
Since we restrict ourselves to the algebra on-shell, 
we substitute the solution $A^{(+)}_{ij \mu}$ of (\ref{solA+}) 
into (\ref{eq-RS}) and (\ref{eq-tRS}). 
Then we obtain 
\ba
\A \A \epsilon^{\mu \nu \rho \sigma} \gamma_{\rho} 
      D_{\sigma} \psi_{R \nu} = 0, 
\label{eq-RS1} \\
\A \A \epsilon^{\mu \nu \rho \sigma} \gamma_{\rho} D_{\sigma} 
      \tilde \psi_{L \nu} - M^{\mu} = 0, 
\label{eq-tRS1}
\ea
where 
\be
M^{\mu} := -{1 \over 2} 
\epsilon^{\mu \nu \rho \sigma} {T^i}_{\rho \sigma} \gamma_i 
\tilde \psi_{L \nu}. 
\label{MM}
\ee
The $M^{\mu}$ of (\ref{MM}) vanishes for $N = 1$ SUGRA 
because of a Fierz transformation. 
After the contraction of (\ref{eq-RS1}) and (\ref{eq-tRS1}) 
with $\gamma_{\mu}$, we have 
\ba
\A \A S^{\mu \nu} D_{\mu} \psi_{R \nu} = 0, \\
\A \A S^{\mu \nu} D_{\mu} \tilde \psi_{L \nu} 
      + {1 \over 4} \gamma_{\mu} M^{\mu} = 0. 
\ea
Further using the relation 
$2 \gamma^{\lambda} S^{\mu \nu} = -i (g^{\lambda \mu} \gamma^{\nu} 
- g^{\lambda \nu} \gamma^{\mu}) 
+ \epsilon^{\lambda \mu \nu \rho} \gamma_5 \gamma_{\rho}$, 
(\ref{eq-RS1}) and (\ref{eq-tRS1}) become 
\ba
\A \A i \gamma^{\rho} 
      (D_{\rho} \psi_{R \lambda} 
                  - D_{\lambda} \psi_{R \rho}) = 0, 
\label{eq-RS2} \\
\A \A i \gamma^{\rho} 
      (D_{\rho} \tilde \psi_{L \lambda} 
                  - D_{\lambda} \tilde \psi_{L \rho}) 
      + M_{\lambda} + {1 \over 2} \gamma_{\lambda}(\gamma \cdot M) 
      = 0. 
\label{eq-tRS2}
\ea

The commutator algebra 
of the SUSY transformations in ${\cal L}^{(+)}$ 
on the complex tetrad is easily calculated as 
\ba
\A \A [\delta_{R1}, \delta_{R2}] e_{\mu}^i = 0 
= [\delta_{L1}, \delta_{L2}] e_{\mu}^i, \\
\A \A [\delta_{R1}, \delta_{L2}] e_{\mu}^i 
= 2i D_{\mu} (\overline{\tilde \alpha}_{2R} \gamma^i \alpha_{1R}), 
\label{com-eRL}
\ea
where the equation for the self-dual connection 
is used in (\ref{com-eRL}). 
Therefore, we have 
\be
[\delta_1, \delta_2] e_{\mu}^i 
= D_{\mu} (\xi^i + \eta^i), 
\label{com-e}
\ee
where $\delta := \delta_R + \delta_L$ and 
\ba
\A \A \xi^i := 2i \ \overline{\tilde \alpha}_{2R} \gamma^i \alpha_{1R}, \\
\A \A \eta^i := 2i \ \overline \alpha_{2L} \gamma^i \tilde \alpha_{1L}. 
\ea

Next we calculate the commutator algebra 
of the SUSY transformations in ${\cal L}^{(+)}$ 
on $\psi_{R \mu}$ and $\tilde \psi_{L \mu}$. 
For $\psi_{R \mu}$, we have 
\ba
\A \A [\delta_{R1}, \delta_{R2}] \psi_{R \mu} = 0 
= [\delta_{L1}, \delta_{L2}] \psi_{R \mu}, \\
\A \A [\delta_{R1}, \delta_{L2}] \psi_{R \mu} 
= -i (\delta_{L2} A^{(+)}_{ij \mu}) S^{ij} \alpha_{1R}. 
\label{com-pR}
\ea
Further we rewrite (\ref{com-pR}) as follows. 
Using the identity 
\be
\eta_{ij} \epsilon_{klmn} = \eta_{ik} \epsilon_{jlmn} 
+ \eta_{il} \epsilon_{kjmn} + \eta_{im} \epsilon_{kljn} 
+ \eta_{in} \epsilon_{klmj} 
\ee
stated in \cite{FN}, we can rewrite (\ref{tpsi-A+}) as 
\ba
\delta_L A^{(+)}_{ij \mu} \A = \A 
{\rm self \! \! - \! \! dual \ part \ of} \nonu
\A \A \{ - {\epsilon_{ij}}^{\lambda \rho} \overline{\tilde \alpha}_R 
\gamma_{\lambda}(D^{(+)}_{\mu} \psi_{R \rho} 
- D^{(+)}_{\rho} \psi_{R \mu}) 
+ e_{\mu [i} {\epsilon_{j]}}^{\lambda \rho \sigma} 
\overline{\tilde \alpha}_R \gamma_{\lambda} D^{(+)}_{\rho} 
\psi_{R \sigma} \}. 
\label{tdelA}
\ea
The last term in (\ref{tdelA}) vanishes by means 
of the field equation (\ref{eq-RS}), 
and hence it can be omitted. 
Substituting the first term of (\ref{tdelA}) 
into (\ref{com-pR}), and using the solution 
$A^{(+)}_{ij \mu}$ of (\ref{solA+}), we get 
\be
[\delta_{R1}, \delta_{L2}] \psi_{R \mu} 
= \xi^{\rho} 
(D_{\rho} \psi_{R \mu} - D_{\mu} \psi_{R \rho}) 
+{1 \over 4} (\xi \cdot \gamma) \gamma^{\rho} 
(D_{\rho} \psi_{R \mu} - D_{\mu} \psi_{R \rho}) 
\label{com-pR1}
\ee
after a Fierz transformation. 
The last term in (\ref{com-pR1}) may be dropped 
because of the field equation of (\ref{eq-RS2}). 
Finally, we obtain 
\be
[\delta_1, \delta_2] \psi_{R \mu} 
= (\xi^{\rho} + \eta^{\rho}) 
(D_{\rho} \psi_{R \mu} - D_{\mu} \psi_{R \rho}), 
\label{com-pRtot}
\ee
discarding those terms which vanish due to 
the matter field equation. 

In the same way, we have the following algebra 
on $\tilde \psi_{L \mu}$ by using (\ref{deltA-}), 
(\ref{tdeltA-}) and the field equation of (\ref{eq-tRS1}): 
\ba
\A \A [\delta_{R1}, \delta_{R2}] \tilde \psi_{L \mu} = 0 
= [\delta_{L1}, \delta_{L2}] \tilde \psi_{L \mu}, \\
\A \A [\delta_{R1}, \delta_{L2}] \tilde \psi_{L \mu} 
= \xi^{\rho} 
(D_{\rho} \tilde \psi_{L \mu} - D_{\mu} \tilde \psi_{L \rho}) 
+ {1 \over 4} (\xi \cdot \gamma) \gamma^{\rho} 
(D_{\rho} \tilde \psi_{L \mu} - D_{\mu} \tilde \psi_{L \rho}). 
\label{com-tpL}
\ea
The last term in (\ref{com-tpL}) may be dropped 
due to the field equation of (\ref{eq-tRS2}) and we find 
\be
[\delta_1, \delta_2] \tilde \psi_{L \mu} 
= (\xi^{\rho} + \eta^{\rho}) 
(D_{\rho} \tilde \psi_{L \mu} - D_{\mu} \tilde \psi_{L \rho}). 
\label{com-tpLtot}
\ee
If the tetrad is real and 
$\overline{\tilde \psi}_{R \mu} = \overline \psi_{R \mu}$, 
the algebra of (\ref{com-e}), (\ref{com-pRtot}) 
and (\ref{com-tpLtot}) coincides with 
that of the usual $N = 1$ SUGRA \cite{FN}. 

The commutator algebra of (\ref{com-e}), (\ref{com-pRtot}) 
and (\ref{com-tpLtot}) can be interpreted 
as the complex extension of the usual $N = 1$ SUGRA. 
Indeed, the infinitesimal general coordinate transformations 
of the fields can be written as \cite{FN} 
\ba
\A \A \delta_G e_{\mu}^i = D_{\mu} \epsilon^i 
- \epsilon^{\rho} {A^i}_{j \rho} e_{\mu}^j 
+ \epsilon^{\rho} {T^i}_{\mu \rho}, \\
\A \A \delta_G \psi_{R \mu} = 
\epsilon^{\rho} (D_{\rho} \psi_{R \mu} - D_{\mu} \psi_{R \rho}) 
- {i \over 2} \epsilon^{\rho} A_{ij \rho} S^{ij} \psi_{R \mu} 
+ D_{\mu}(\epsilon \cdot \psi_R), \\
\A \A \delta_G \tilde \psi_{L \mu} = 
\epsilon^{\rho} (D_{\rho} \tilde \psi_{L \mu} 
                       - D_{\mu} \tilde \psi_{L \rho}) 
- {i \over 2} \epsilon^{\rho} A_{ij \rho} S^{ij} 
  \tilde \psi_{L \mu} 
+ D_{\mu}(\epsilon \cdot \tilde \psi_L) 
\ea
with $\epsilon^{\mu}$ being the infinitesimal 
transformation parameter: $x'^{\mu} = x^{\mu} - \epsilon^{\mu}$. 
If we take $\epsilon^{\mu}$ as 
\be
\epsilon^{\mu} := \xi^{\mu} + \eta^{\mu}, 
\ee
then the commutator algebra of (\ref{com-e}), (\ref{com-pRtot}) 
and (\ref{com-tpLtot}) can be rewritten as follows: 
\ba
\A \A [\delta_1, \delta_2] e_{\mu}^i 
      = \delta_G e_{\mu}^i + \epsilon^{\rho} {A^i}_{j \rho} e_{\mu}^j 
      - \epsilon^{\rho} {T^i}_{\mu \rho}, 
\label{com-egen} \\
\A \A [\delta_1, \delta_2] \psi_{R \mu} 
      = \delta_G \psi_{R \mu} 
      + {i \over 2} \epsilon^{\rho} A_{ij \rho} S^{ij} \psi_{R \mu} 
      - D_{\mu}(\epsilon \cdot \psi_R) 
\label{com-pRgen} \\
\A \A [\delta_1, \delta_2] \tilde \psi_{L \mu} 
      = \delta_G \tilde \psi_{L \mu} 
      + {i \over 2} \epsilon^{\rho} A_{ij \rho} S^{ij} \tilde \psi_{L \mu} 
      - D_{\mu}(\epsilon \cdot \tilde \psi_L). 
\label{com-tpLgen}
\ea
The first and second terms of (\ref{com-egen}) to (\ref{com-tpLgen}) 
describe the complex general coordinate transformations 
and the field-dependent 
complex local Lorentz transformations, respectively. 
The third terms in (\ref{com-pRgen}) and (\ref{com-tpLgen}) 
are just the right- and left-handed SUSY 
transformations generated by the field-dependent parameters 
$\alpha'_R := -(1/2) \epsilon \cdot \psi_R$ 
and $\tilde \alpha'_L := -(1/2) \epsilon \cdot \tilde \psi_L$. 
Similarly, the third term in (\ref{com-egen}) 
can be interpreted as the sum of the right- and left-handed SUSY 
transformations by using (\ref{torsion}): 
\ba
- \epsilon^{\rho} {T^i}_{\mu \rho} 
\A = \A i \epsilon^{\rho} 
     \overline{\tilde \psi}_{R [\mu} \gamma^i \psi_{R \rho]} \nonu
\A = \A -i (\overline{\tilde \psi}_{R \mu} \gamma^i \alpha'_R 
        + \overline \psi_{L \mu} \gamma^i \tilde \alpha'_L). 
\ea
Note that the structure constants defined by these results 
are field-dependent, i.e., structure functions, 
as in the case of the usual $N = 1$ SUGRA. 

So far we have considered the SUSY transformations 
in ${\cal L}^{(+)}$ of (\ref{L+}) with the complex tetrad. 
In the usual local field theory, 
the spinor fields $\psi$ and $\overline \psi$ 
are not independent of each other. 
However, since the independent 
$\psi_{R \mu}$ and $\overline{\tilde \psi}_{R \mu}$ 
have been used in ${\cal L}^{(+)}$, 
it is difficult to see 
the correspondence of ${\cal L}^{(+)}$ 
to the usual local field theory. 
This correspondence can be seen 
if we add the complex conjugate of the chiral Lagrangian 
density, $\overline{{\cal L}^{(+)}}$, 
to the Lagrangian density ${\cal L}^{(+)}$. 
For example, in special relativistic limit, 
the total Lagrangian density of (Majorana) Rarita-Schwinger fields, 
${\cal L}_{RS}^{tot} := 
{\cal L}_{RS}^{(+)} + \overline{{\cal L}_{RS}^{(+)}}$, 
becomes 
\be
L^{tot}_{RS} = \epsilon^{\mu \nu \rho \sigma} 
               \overline {\tilde \psi}_{\mu} \gamma_5 \gamma_\rho 
               \partial_\sigma \psi_{\nu}, \label{LRStots}
\ee
which can be diagonalized as 
\be
L^{tot}_{RS} = \epsilon^{\mu \nu \rho \sigma} 
              (\overline \psi_{\mu}^1 \gamma_5 \gamma_\rho 
               \partial_\sigma \psi_{\nu}^1 
             - \overline \psi_{\mu}^2 \gamma_5 \gamma_\rho 
               \partial_\sigma \psi_{\nu}^2), 
\label{LRStots}
\ee
with $\psi_{\mu}^1 := (1/2)(\psi_{\mu} + {\tilde \psi}_{\mu})$ 
and $\psi_{\mu}^2 := (1/2)(\psi_{\mu} - {\tilde \psi}_{\mu})$. 
The minus sign in (\ref{LRStots}) means 
the appearance of negative energy states. 
The SUSY transformations in the total Lagrangian density, 
${\cal L}^{tot} := {\cal L}^{(+)} + \overline{{\cal L}^{(+)}}$, 
can be constructed as in ${\cal L}^{(+)}$. 

We have also seen that the SUSY algebra 
for $N = 1$ chiral SUGRA with the complex tetrad 
closes only on-shell, 
because the field equation terms appear 
in the SUSY algebra on (Majorana) Rarita-Schwinger fields. 
In the usual $N = 1$ SUGRA, additional auxilialy fields 
are introduced in order that the SUSY algebra 
closes off-shell \cite{SWFN}. 
We are now trying to construct 
such a formulation for $N = 1$ chiral SUGRA.

We would like to thank the members of Physics Department 
at Saitama University 
for discussions and encouragement.



\end{document}